\documentclass[prl,reprint,amssymb,superscriptaddress]{revtex4-2}

\usepackage{xcolor}

\usepackage{bm}

\usepackage{amsmath}  
\usepackage{amsfonts}
\usepackage{graphicx}
\usepackage{hyperref}
\usepackage{comment}
\usepackage[normalem]{ulem}
\hypersetup{
	colorlinks=true,
	citecolor=blue,
	linkcolor=red,
	urlcolor = blue
}

\begin{document}
	
\title{Charge and spin dynamics in spintronic THz emitters from Time-Dependent Density Functional Theory}

\author{Ali Kefayati}
\affiliation{Institute of Physics, Academia Sinica, Taipei 115201, Taiwan}
\email{alikefay@as.edu.tw}

\begin{abstract}
Spintronic terahertz (THz) emitters based on ferromagnet/heavy-metal heterostructures provide efficient broadband sources of THz radiation, yet the microscopic relation between ultrafast spin transport and charge redistribution remains incompletely understood. Here, we employ time-dependent density functional theory to resolve the coupled charge, spin, and magnetization dynamics of Co/Pt and Co/W bilayers in real time and space. Despite qualitatively similar interlayer spin-current dynamics, the two heterostructures exhibit markedly different interlayer charge transfer: Co/Pt shows a relatively small and partially reversed transfer from Pt to Co, whereas Co/W exhibits a larger and persistent transfer from Co to W. We trace these differences to the material-dependent ground-state electronic structure and the spin- and orbital-resolved phase space accessible to photoexcited carriers. Spin-resolved dynamics further reveal coexisting spin-orbit-mediated redistribution and optically induced intersite spin transfer, including a delayed minority-spin transfer from the heavy metal to Co common to both systems. While net electronic spin angular momentum is transferred between the layers, we find no corresponding net transfer of electronic orbital angular momentum. In Co/W, the persistent redistribution of electronic density from interstitial to atom-centered states additionally provides strong evidence for \emph{light-induced electronic localization}. These results demonstrate that ultrafast spin transport does not uniquely determine charge or angular-momentum redistribution in spintronic THz emitters, which instead depends critically on the material-specific electronic phase space.
\end{abstract}

\maketitle

%-------------------------------------------------------------------------------------

\section{Introduction}
\begin{figure}
    \centering
    \includegraphics[width = \linewidth]{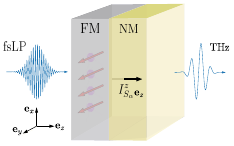}
    \caption{Schematic view of the FM/NM bilayer used as a spintronic THz emitter. A femtosecond laser pulse (fsLP) excites the bilayer, producing ultrafast magnetization, charge, and spin dynamics. Co is used as the FM layer and Pt or W as the NM layer.}
    \label{fig:FigI}
\end{figure}

The discovery of femtosecond laser-induced demagnetization opened a new regime for manipulating magnetic order on time scales much faster than conventional spin-lattice relaxation.\cite{Beaurepaire1996, Beaurepaire2004, Hilton2004}. Since then, extensive experimental and theoretical efforts have sought to identify the microscopic processes governing the redistribution of charge, spin, orbital, and lattice angular momentum following ultrafast optical excitation of magnetic materials \cite{Zhang2000a, Malinowski2008, Zhang2008, Kirilyuk2010, Battiato2010, LaOVorakiat2012, Rudolf2012, Eschenlohr2013, Choi2014, Vodungbo2016, Eschenlohr2017, Hurst2018, Tengdin2018, Huang2019, Huang2020, Buhlmann2020, Scheid2022, VarelaManjarres2024, Chen2025, Li2026Ultrafast}. An important development was the introduction of spintronic THz emitters (STEs), typically consisting of a ferromagnetic (FM) metallic layer adjacent to a heavy nonmagnetic metal (NM)~\cite{kampfrath2013,Seifert2016,Wu2017}. As illustrated schematically in Fig.~\ref{fig:FigI}, excitation by a femtosecond laser pulse produces ultrafast magnetization and current dynamics and generates broadband electromagnetic radiation in the THz frequency range. The conventional interpretation of STE operation is based on laser-induced generation of a spin current in the FM layer followed by its propagation into the NM layer. Strong spin-orbit coupling (SOC) in the NM then converts the injected spin current into an in-plane charge current, commonly described in terms of the inverse spin Hall effect (ISHE)~\cite{kampfrath2013}. This picture has been remarkably useful for interpreting experiments and optimizing STE materials \cite{kampfrath2013, Huisman2015, Seifert2016, Wu2017, Zhang2017b, Freimuth2017, Zhang2020, Gueckstock2021, Liu2021, Wu2021, Wahada2022, Wang2023, Cheng2023, Leitenstorfer2023, Jechumtal2024, Shelukhin2025Enhanced, Song2025Ultrafast, Abdelaziem2025Terahertz, Garcia2025Terahertz, Yang2026Intense, Shrestha2026Crystal, Liu2026HoCo, Ge2026Optically}.
However, an optically excited FM/NM interface constitutes a strongly nonequilibrium and spatially inhomogeneous electronic system. Interface hybridization, proximity effects \cite{Belashchenko2016, MarmolejoTejada2017, Dolui2017, Jungfleisch2018a, Lu2020, Dolui2020a, Gupta2020}, SOC, intersite spin transfer, and direct optical excitation can coexist on femtosecond time scales. A description based solely on equilibrium transport coefficients or a single spin-to-charge conversion mechanism \cite{Battiato2012, Tveten2015, Fognini2017, Pavel2018, Nenno2018, Nenno2019, Tows2019, Beens2020, Lu2021, Beens2022a, Lichtenberg2022, Rouzegar2022, Beens2023, Schmidt2023, Busch2023Nov, Foggetti2025, Weissenhofer2025, Foggetti2026Optimizing} can therefore hide important microscopic processes. 

First-principles TDDFT \cite{Noda2019, Tancogne-Dejean2020, elk} provides a complementary approach because the electronic charge and spin densities can be propagated directly in real time without introducing electronic temperatures or phenomenological electronic relaxation rates. Previous TDDFT studies have established important roles for SOC-mediated spin flips and optically induced intersite spin transfer (OISTR) in ultrafast magnetization dynamics \cite{Krieger2015, Stamenova2016, Krieger2017, Simoni2017, Dewhurst2018, Chen2019a, Chen2019c, Acharya2020, Pellegrini2022, Sharma2022, Wu2024, 
Mrudul2024, Kefayati2026Light}. The microscopic role of interlayer charge redistribution, however, remains less clear. In particular, previous first-principles work has attributed the enhancement of ultrafast demagnetization in FM/NM heterostructures to nonlocal $sp$--$d$ spin-exchange coupling processes that redistribute spin angular momentum without requiring net charge transfer between the constituent layers \cite{chen2023light}. Thus, whether the ultrafast interlayer spin transport generally decoupled from charge transfer, or can the redistribution of spin angular momentum be accompanied by a net transfer of electronic charge between the layers, remains an open question.

In our recent studies, we employed TDDFT together with Jefimenko's equations to examine the microscopic origin of electromagnetic radiation from spintronic THz emitters \cite{Kefayati2024, Kefayati2025Deciphering}. These calculations demonstrated that time-dependent charge currents generated inside both FM and NM layers contribute to the emitted electromagnetic field and that the relation between injected spin current and THz radiation can be considerably more complex than a simple bulk-ISHE picture, while the interlayer spin current remains  proportional to the temporal variation of the magnetization i.e. $I_{S_\alpha}^z \propto d\mathbf{M}/dt$.

These results raise a more fundamental question: how are spin transfer and charge transfer related in an optically driven FM/NM heterostructure, and what determines the redistribution of photoexcited electrons between the two layers? Here we address this question by comparing two prototypical STE heterostructures, Co/Pt and Co/W. The two systems exhibit similar qualitative spin-current dynamics but markedly different charge redistribution and magnetization dynamics. In contrast to the pure-spin-transport picture based on $sp$-$d$ exchange coupling \cite{chen2023light}, we find a finite time-dependent transfer of electronic charge between the FM and NM layers, whose magnitude and even direction depend strongly on the NM material. We resolve the total, layer-resolved, spin-resolved, and spatially resolved electronic dynamics and connect them to the ground-state projected density of states (PDOS). Our results reveal that the electronic phase space available to photoexcited carriers plays a central role in selecting the dominant microscopic pathway. In Co/Pt, excitation predominantly populates spatially nonlocal states and leads to strong demagnetization of Co while Pt remains nearly nonmagnetic. In Co/W, substantial population transfer into atom-centered states accompanies charge and spin transfer between Co and W and induces a persistent W magnetization over the simulated time window. We additionally identify a delayed minority-spin OISTR-like channel from NM to Co that occurs in both heterostructures, while a pronounced majority-spin Co-to-W channel appears in Co/W. Furthermore, the final interstitial electron density in Co/W falls below its initial value while the local-state population increases. We identify this persistent redistribution toward atom-centered states as strong evidence for light-induced electronic localization. Thus, even when similar spin currents are generated, the subsequent charge, spin, and magnetization dynamics can differ qualitatively because of the electronic structure of the NM layer.

%-------------------------------------------------------------------------------------

\section{Computational Method}
TDDFT formalism~\cite{Marques2006} is the time-dependent version of  Kohn-Sham (KS) equation given by (using $\hbar=1$) 
\begin{eqnarray}\label{eq:tddft}
    i\frac{\partial {\bm \psi}_j(\textbf{r},t)}{\partial t} & = & \bigg [ \frac{1}{2m_e} \bigg(-i\nabla+\frac{1}{c}\mathbf{A}_\mathrm{ext}(t) \bigg)^2 + v_s(\textbf{r},t) \nonumber\\
    & +& \frac{1}{2c} {\bm \sigma} \cdot \textbf{B}_s(\textbf{r},t)\nonumber\\
    & + &\frac{1}{4c^2}{\bm \sigma} \cdot \big( \nabla v_s(\textbf{r},t)\times -i\nabla\big) \bigg]\psi_j(\textbf{r},t),
\end{eqnarray}
where ${\bm \psi}_j(\textbf{r},t)$ are two-component Pauli spinors of KS quasiparticle with quantum number $j$; $m_e$ is the free electron mass; $\mathbf{A}_\mathrm{ext}(t)$ is the vector potential of the applied external excitation; $v_s(\textbf{r},t) = v_{\mathrm{ext}}(\textbf{r},t) + v_{\mathrm{H}}(\textbf{r},t) + v_{\mathrm{XC}}(\textbf{r},t) $ is the effective KS potential, as the sum of the external potential $v_{\mathrm{ext}}$ from the atomic nuclei, the Hartree potential $v_{\mathrm{H}}$, and the exchange-correlation (XC) potential $v_\mathrm{XC}$;   $\mathbf{B}_s(\textbf{r},t)=\mathbf{B}_\mathrm{ext}(t)+\mathbf{B}_\mathrm{XC}(t)$ is the KS magnetic field with $\mathbf{B}_\mathrm{ext}$ being the external magnetic field ($\mathbf{B}_\mathrm{ext}=0$ in our calculation), and $\mathbf{B}_\mathrm{XC}$ the XC magnetic field; \mbox{${\bm \sigma}=(\sigma_x, \sigma_y, \sigma_z)$} is the vector of the Pauli matrices; and the last term accounts SOC. 

Solving the TDDFT equation in Eq.\ref{eq:tddft}, the Pauli spinors are used to calculated the charge, current, and magnetization densities. 
The charge and magnetization densities are
\begin{align}
n(\mathbf{r},t)&=\sum_j \psi_j^\dagger(\mathbf{r},t) \psi_j(\mathbf{r},t) \\
\mathbf{m}(\mathbf{r},t)&=\sum_j \psi_j^\dagger(\mathbf{r},t) {\bm \sigma} \psi_j(\mathbf{r},t)
\end{align}
with the total magnetization,
\begin{equation}
\mathbf{M}(t)=\int\!\! d^3 r\, \mathbf{m}(\mathbf{r},t).
\end{equation}
 
The microscopic charge current density is given by 
\begin{equation}
\mathbf j(\mathbf r,t)=-\frac{e}{m}\sum_{j}^{\rm occ}\operatorname{Re}
\left[\psi_{j}^\dagger\left(-i\hbar\nabla+e\mathbf A_\mathrm{ext}(t)\right)\psi_{j}\right] 
\end{equation}
and the macroscopic charge current density can be obtained as 
\begin{equation}
 \mathbf J(t)
=\frac{1}{\Omega}\int_\Omega d^3r\,\mathbf j(\mathbf r,t). 
\end{equation} 
The layer-resolved current densities are obtained by integrating the microscopic current density over the corresponding spatial regions. 

The calculations are performed within adiabatic local density approximation (ALDA) for XC functional ~\cite{Lacombe2023} using the full-potential linearized augmented plane-wave (LAPW) method as implemented in the ELK code~\cite{Dewhurst2016,elk}. The grid of $\mathbf{k}$ vectors is chosen as $7 \times 7$ for both Co/Pt or Co/W. The bilayers are in the $xy$-plain and $z$ direction is the out-of-plane direction. The equilibrium Co magnetization is oriented along the $y$ direction. After obtaining the GS, the systems are excited by $x$-polarized Gaussian laser pulse with central wavelength \mbox{$800$ nm}, FWHM $\simeq 25$ fs, the peak intensity of  $2.42$ TW/cm$^2$, and 22 mJ/cm$^2$ fluence. Since the wavelength of the optical pulse much larger than the supercell, we assume dipole approximation and disregard spatial dependence of $\mathbf{A}_\mathrm{ext}(t)$. All time-dependent (TD) outputs presented here are smoothed using a Gaussian filter with a standard deviation of approximately $6.05$ fs (FWHM $\approx$ 14.2 fs). Unless otherwise stated, atomic units are used throughout this paper. We ignore the dynamics of the lattice nuclei since those dynamics are not relevant in the first 100 fs of the excitation \cite{Tauchert2022, Wu2024, Mrudul2025Generation}.

%-------------------------------------------------------------------------------------
\subsection{Local and nonlocal electronic populations} 
Within the LAPW basis, real space is partitioned into atom-centered muffin-tin (MT) regions and the interstitial region (IR) between them. We refer to electronic density within the muffin-tin regions ($n_{\text{MT}}$) as the \emph{local electronic population}, whereas density in the interstitial region ($n_{\text{IR}}$) is referred to as the \emph{nonlocal electronic population}. These terms therefore provide an operational measure of redistribution of electronic density between atom-centered and spatially extended regions of the heterostructure.

Because SOC is included, spin is not a conserved quantum number. We nevertheless resolve the instantaneous spin density by projection along the equilibrium magnetization direction $+\hat y$ and $-\hat y$, denoted as majority ($\uparrow$) and minority ($\downarrow$) channels, respectively.

%-------------------------------------------------------------------------------------

\section{Results}

\begin{figure*}[t]
    \centering
    \includegraphics[width = \linewidth]{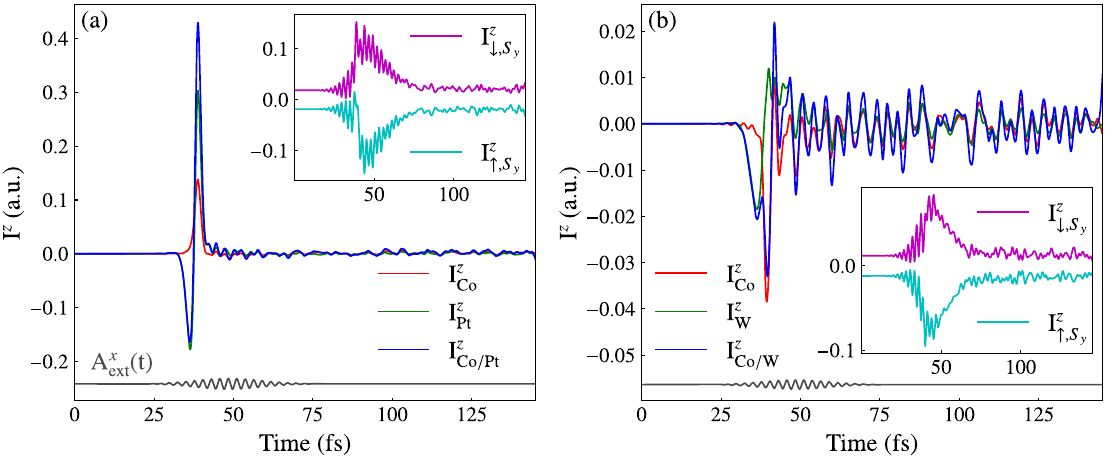}
    \caption{The out-of-plane total and layer-resolved charge current for the (a) Co/Pt, and (b) Co/W bilayers. The inset shows the spin current in each spin channel. The total charge current is the sum of the FM and the NM layers. The vector potential of the laser pulse ($\mathrm{A}^x_{\mathrm{ext}}(t)$) is shown in gray.}
    \label{fig:FigII}
\end{figure*}

Figure~\ref{fig:FigII} shows the total and layer-resolved out-of-plane charge current densities in Co/Pt and Co/W. The total current densities satisfy
\begin{equation}
I^z_{\rm Co/Pt}=I^z_{\rm Co}+I^z_{\rm Pt},\qquad
I^z_{\rm Co/W}=I^z_{\rm Co}+I^z_{\rm W}.
\end{equation}
There is a striking difference in the $I^z_{\text{Co}}$ of the two bilayers. $I^z_{\text{Co}}$ exhibits a positive peak in Co/Pt but a negative peak in Co/W. Together with the layer-resolved charge redistribution discussed below, this indicates qualitatively different interlayer charge-transfer dynamics in the two heterostructures. This also indicates a net electronic charge transfer between FM and NM layers beyond merely a nonlocal exchange coupling \cite{chen2023light}. Particularly revealing is the early excitation regime, where the net Co charge current remains approximately zero even though the charge current in the NM layer as well as the total charge current are nonzero. In addition, the spin-resolved currents shown in the insets of Fig.~\ref{fig:FigII} are individually substantial. During this interval,
\begin{equation}
I_{\rm Co}^{z,\uparrow}\simeq-I_{\rm Co}^{z,\downarrow},
\end{equation}
and consequently
\begin{equation}
I_{\rm Co}^{z}=I_{\rm Co}^{z,\uparrow}+I_{\rm Co}^{z,\downarrow}\simeq0,
\end{equation}
whereas the corresponding spin current,
\begin{equation}
I_{\rm Co}^{z,s}\propto I_{\rm Co}^{z,\uparrow}-I_{\rm Co}^{z,\downarrow},
\end{equation}
remains finite \cite{Kefayati2025Deciphering}. Thus, the absence of a net charge current during the early stage of excitation does not imply the absence of electronic transport. Instead, oppositely directed contributions from the two spin channels approximately cancel in charge while adding in spin. The simultaneous spin-resolved density dynamics discussed below are consistent with strong SOC-mediated spin mixing during this early-time regime.

\begin{figure*}[t]
    \centering
    \includegraphics[width = \linewidth]{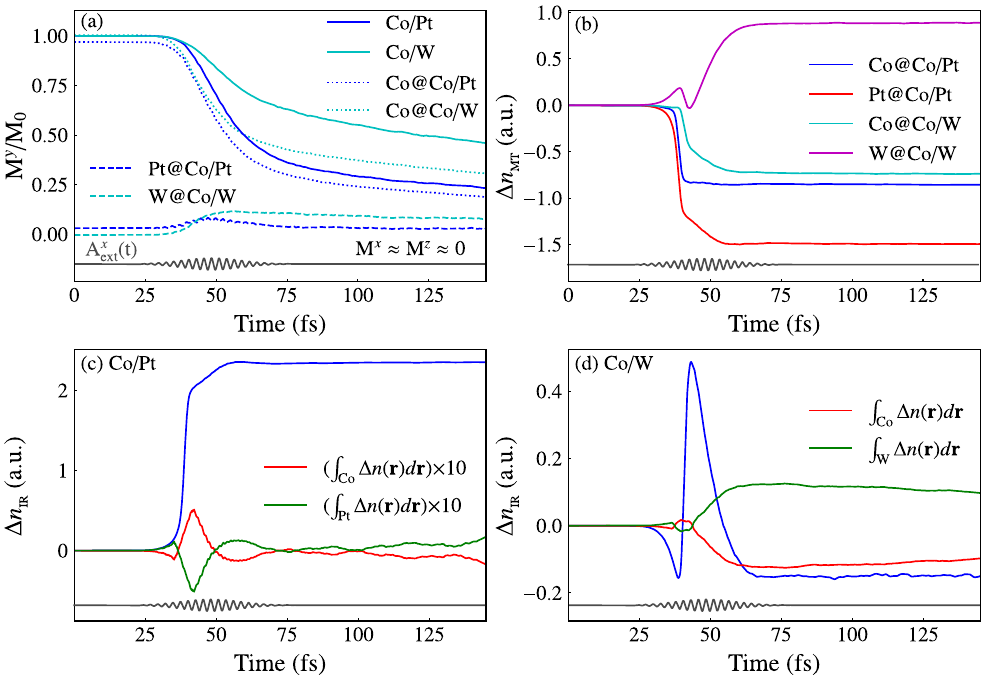}
    \caption{(a) Total and layer-resolved magnetization dynamics of Co/Pt and Co/W. (b) Dynamics of the local electronic populations. (c,d) Nonlocal and layer-resolved excess charge densities of Co/Pt and Co/W, respectively. In Co/Pt the nonlocal density remains enhanced, whereas in Co/W it eventually falls below its initial value.}
    \label{fig:FigIII}
\end{figure*}

Fig. \ref{fig:FigIII} (a) compares the magnetization dynamics of Co/Pt and Co/W. Although both systems demagnetize rapidly after the optical excitation, Co/Pt exhibits faster and quantitatively stronger. The layer-resolved magnetization reveals a qualitatively different role of the heavy-metal layer. Pt remains nonmagnetic, whereas W develops an appreciable persistent induced magnetization \cite{Dewhurst2018Sub}. The smaller net demagnetization of Co/W is partly because of the magnetization of the W layer and partly due to the relatively large magnetization of the nonlocal electronic population as will be discussed below. This indicates that the two NM layers
do not act merely as passive reservoirs receiving equivalent spin currents; their distinct electronic structures participate
differently in the nonequilibrium dynamics.

The local electronic populations also respond differently as depicted in Fig. \ref{fig:FigIII} (b). In Co/Pt, the local populations decrease in
both layers, whereas the W local population increases
in Co/W. The nonlocal electronic populations provides more evidence for the different responses of Pt and W. In Co/Pt, the nonlocal electronic population increases strongly and remains enhanced after the excitation, demonstrating substantial transfer of electronic weight from atom-centered
states into spatially extended states, Fig. \ref{fig:FigIII} (c). Co/W behaves qualitatively differently, Fig \ref{fig:FigIII} (d). Its nonlocal electronic population increases after a small initial reduction during optical excitation but subsequently decreases and, most importantly, becomes smaller than its equilibrium value after the pulse,
\begin{equation}
\Delta n_{\rm IR}(t_f)<0.
\label{eq:localization}
\end{equation} 
The layer-resolved excess charge densities depicted in Fig. \ref{fig:FigIII} (c) and (d) for Co/Pt and Co/W, respectively, shows the dynamics of charge transfer and redistribution between Co and the NM layers. This is the central finding of this paper. In Co/Pt, an initial charge transfer from Pt toward Co is relatively small and is subsequently partially reversed. In contrast, Co/W exhibits a substantially larger charge transfer from Co toward W, which does not undergo the corresponding reversal on the simulated time scale. Thus, despite qualitatively similar spin-current dynamics, Co/Pt and Co/W exhibit fundamentally different interlayer charge-transfer dynamics. This behavior points to the available electronic phase space, rather than the spin current alone, as the determining factor. As discussed below, the GS-PDOS shows markedly different laser-accessible final-state phase space in Pt and W, providing a microscopic basis for the contrasting charge-transfer pathways.

\begin{figure*}[t]
    \centering
    \includegraphics[width = \linewidth]{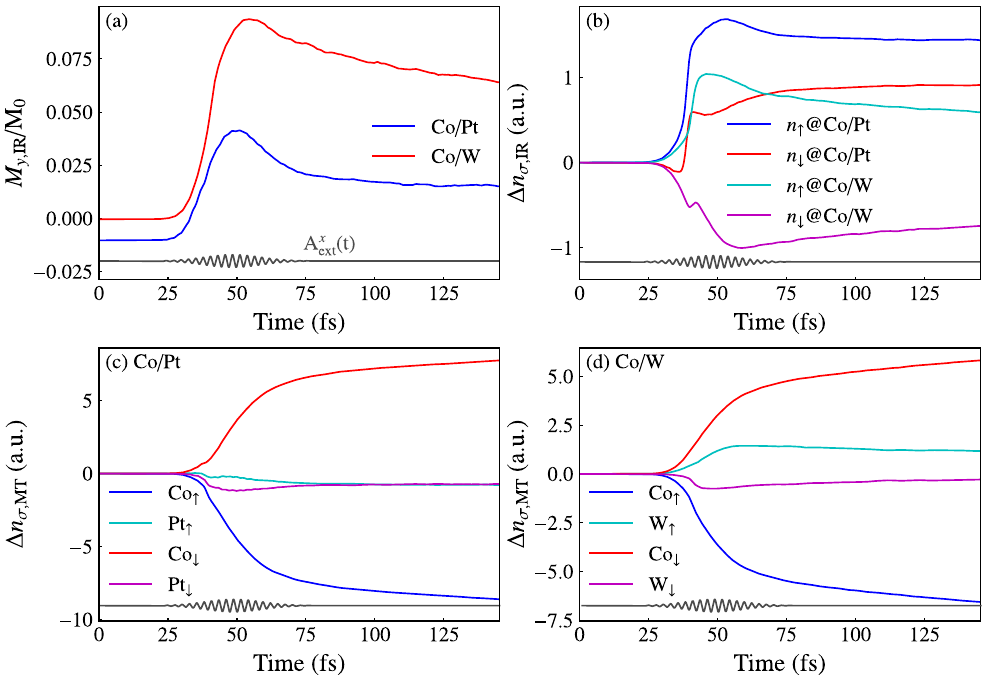}
    \caption{(a) Magnetization associated with the nonlocal electronic population. (b) Spin-resolved nonlocal charge densities. The minority spin density reduction in W indicate a transition to the atom-central local electronic states resulting in a large magnetization of the nonlocal electronic population. (c,d) Spin-resolved local charge densities of Co/Pt and Co/W. The delayed reduction of the NM minority-spin local population, together with the increase of the Co minority-spin population, is consistent with a common ${\rm NM}_{\downarrow}\rightarrow{\rm Co}_{\downarrow}$ OISTR channel. Co/W additionally exhibits a pronounced majority-spin Co-to-W transfer channel.}
    \label{fig:FigIV}
\end{figure*}

To identify the microscopic origin of these differences, Fig.~\ref{fig:FigIV} decomposes the local and nonlocal populations into their spin-resolved contributions. Despite the overall demagnetization, the magnetization associated with the nonlocal electronic population increases in both systems \cite{Krieger2015, Elliott2020} as depicted in Fig. \ref{fig:FigIV} (a). The origin of this behavior becomes clearer from the spin-resolved densities. In Co/Pt, both spin channels contribute to the increased nonlocal population shown is Fig. \ref{fig:FigIV} (b), consistent with optical excitation from predominantly atom-centered states into spatially extended electronic states. Simultaneously, the local Co majority-spin population decreases while the minority-spin population increases, Fig. \ref{fig:FigIV} (c). This behavior is consistent with strong SOC-mediated spin mixing in Co. 

An additional feature, common to both Co/Pt and Co/W, appears in the local minority-spin channel. Following the initial response of the Co layer, the local minority-spin population of the NM layer decreases with a delay, while the corresponding Co minority-spin population increases, see Fig. \ref{fig:FigIV} (c) and (d). The similar temporal pattern in Pt and W suggests a common microscopic channel. We interpret this correlated, delayed response as being consistent with OISTR of local minority-spin electronic population from the NM layer toward available minority-spin states in Co,
\begin{equation*}
{\rm NM}_{\downarrow}\xrightarrow{\text{OISTR}} {\rm Co}_{\downarrow}.
\label{eq:oistrdown}
\end{equation*}
Because this transfer preserves the spin projection, it does not require a spin flip \cite{Dewhurst2018, Siegrist2019}. Its occurrence in both heterostructures is naturally supported by the substantial unoccupied Co minority-spin $d$-state phase space above the Fermi level. Once minority-spin electrons in the NM layer are optically excited, these empty Co states provide accessible final states across the interface.

The Co/W dynamics contain an additional and oppositely directed spin-conserving channel. The local W majority-spin population increases while the Co majority-spin population is depleted, consistent with OISTR from Co toward W,
\begin{equation*}
{\rm Co}_{\uparrow}\xrightarrow{\text{OISTR}} {\rm W}_{\uparrow}.
\label{eq:oistrup}
\end{equation*}
This process contributes to the persistent induced magnetization of W. Thus, OISTR in the present systems is not characterized by a unique direction of electronic transfer. Instead, its direction and spin character are selected by the occupied initial states and the available final-state phase space.

The resulting microscopic picture is therefore more general than a simple division in which Co/Pt is governed by SOC whereas Co/W is governed by OISTR. SOC-mediated spin redistribution and OISTR coexist in both systems. What differs is the relative weight of the available spin-resolved transfer channels. In both bilayers, early SOC-mediated redistribution occurs within Co and a smaller delayed ${\rm NM}_{\downarrow}\rightarrow{\rm Co}_{\downarrow}$ OISTR-like channel is present. In Co/W, the additional ${\rm Co}_{\uparrow}\rightarrow{\rm W}_{\uparrow}$ channel becomes prominent and contributes to the distinctive W magnetization and charge redistribution.

An additional feature emerges by comparing the total charge dynamics in Fig. \ref{fig:FigIII}  (b) with the spin-resolved populations in Figs.\ref{fig:FigIV} (c) and (d). The two observables exhibit distinct temporal offsets between the FM and NM layers. Whereas the total charge population shows a delayed response in the FM relative to the NM, the spin-resolved dynamics display the opposite ordering, with changes in the NM populations developing after those in Co. This indicates that charge and spin populations do not evolve as a single rigid interlayer transfer process. Such behavior is possible because charge and spin probe, respectively, the sum and difference of the two spin-channel populations, \(n=n_\uparrow+n_\downarrow\) and \(m\propto n_\uparrow-n_\downarrow\), so that redistribution between spin channels can produce substantially different temporal signatures in the charge and magnetic observables.

\begin{figure*}[t]
    \centering
    \includegraphics[width = \linewidth]{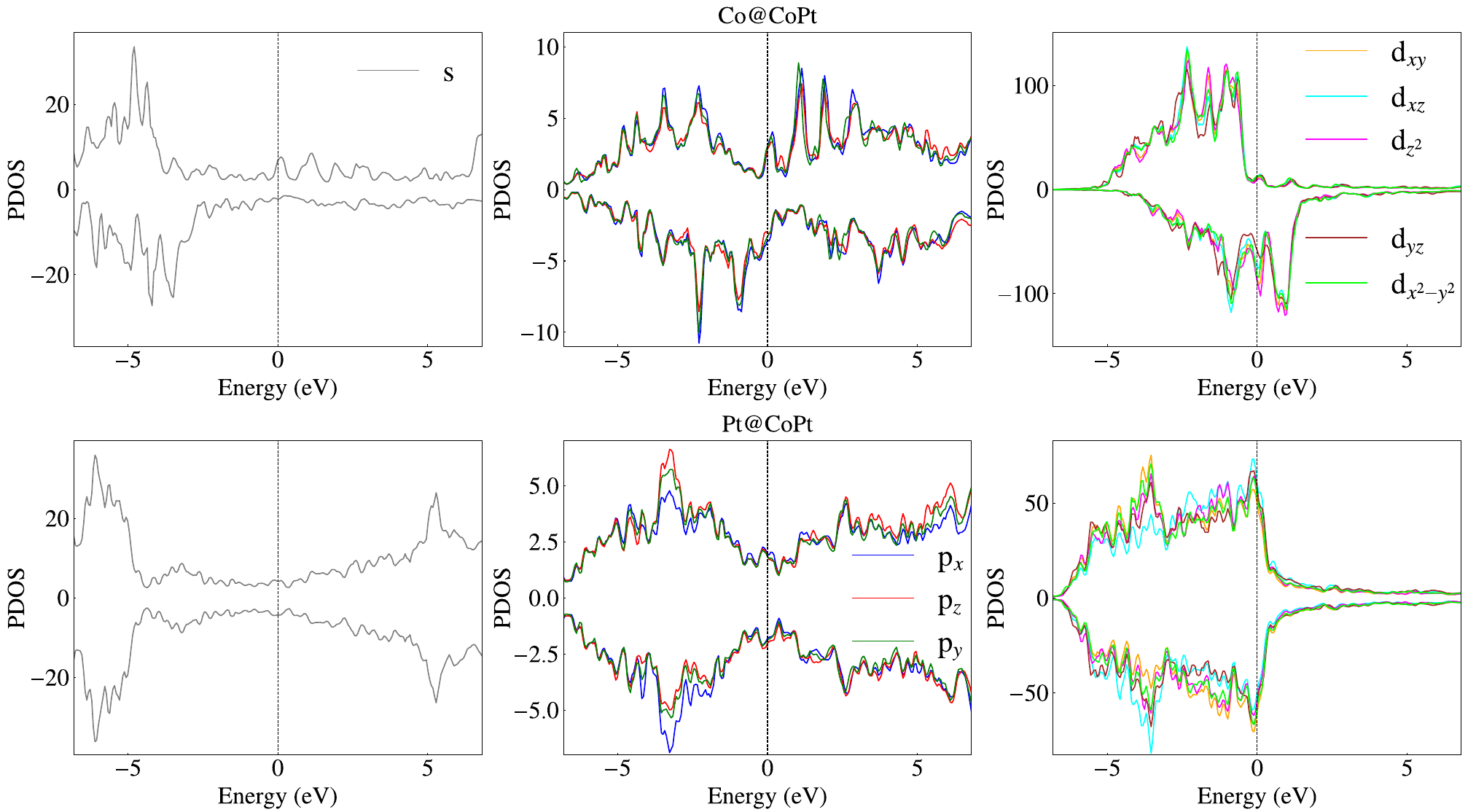}
    \caption{Projected density of states of the Co/Pt bilayer resolved into atomic orbital and spin character.}
    \label{fig:PDOSCoPt}
\end{figure*}

\begin{figure*}[t]
    \centering
    \includegraphics[width = \linewidth]{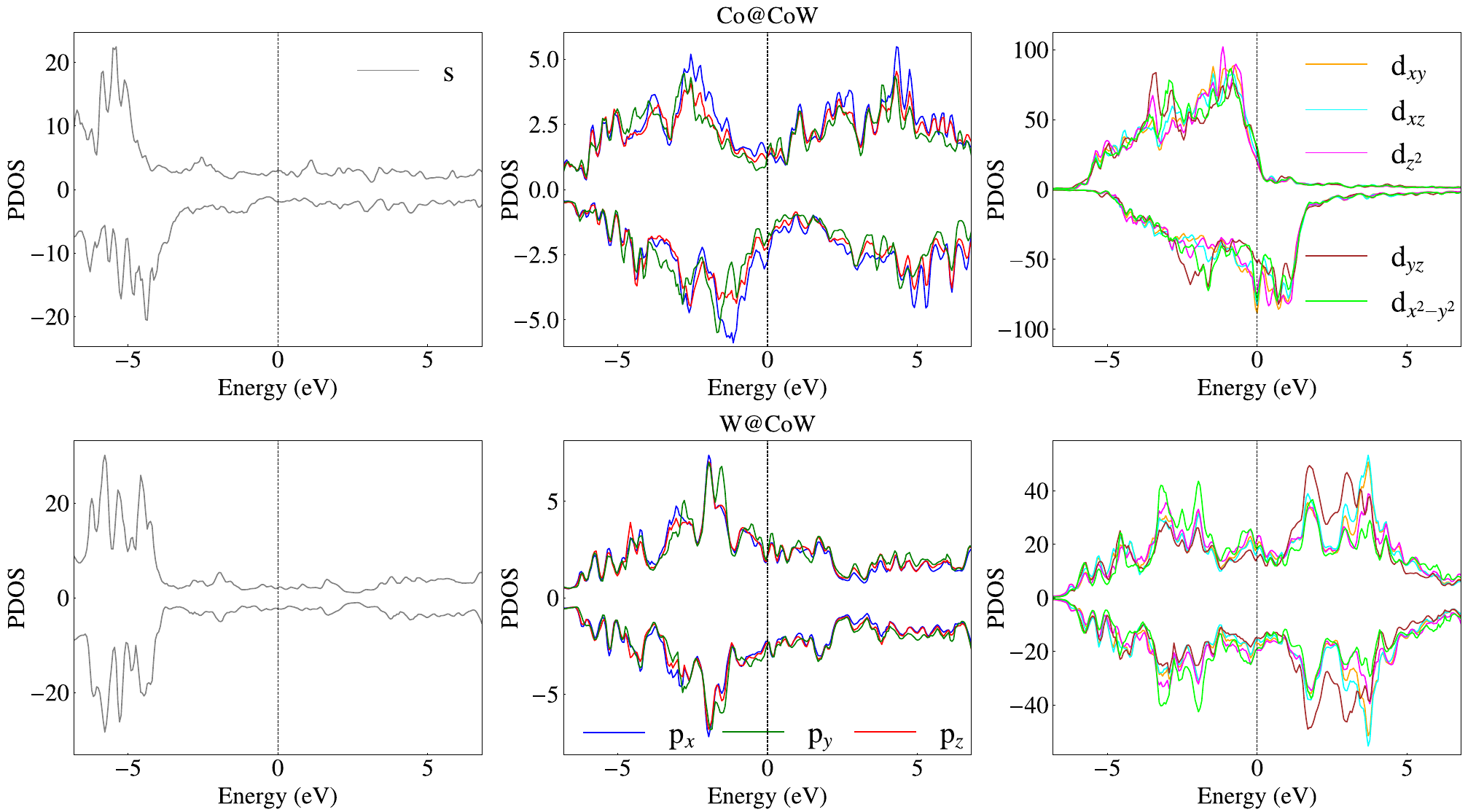}
    \caption{Projected density of states of the Co/W bilayer resolved into atomic orbital and spin character.}
    \label{fig:PDOSCoW}
\end{figure*}

Figures~\ref{fig:PDOSCoPt} and~\ref{fig:PDOSCoW} show the orbital-resolved PDOS of Co/Pt and Co/W, respectively. These ground-state electronic structures provide a microscopic basis for understanding the markedly different nonequilibrium responses. The laser is centered at approximately 1.56~eV. The relevant excitation pathways are therefore governed by occupied and unoccupied states lying within the spectral window accessible to the optical pulse rather than by the DOS at the Fermi level alone.

In Co/Pt, substantial $s$- and $p$-character electronic states are available within the relevant excitation window, Fig. \ref{fig:PDOSCoPt}. Because these orbitals are spatially more extended than strongly atom-centered $d$ orbitals, their availability is consistent with the pronounced increase of interstitial electronic density observed in Fig.~\ref{fig:FigIII} (c). The Co minority-spin $d$ states above the Fermi level simultaneously provide phase space for SOC-mediated redistribution and for the delayed minority-spin transfer from Pt toward Co identified in Fig.~\ref{fig:FigIV} (c).

The situation in Co/W is markedly different, Fig. \ref{fig:PDOSCoW}. A substantial density of W $d$ states is available above the Fermi level, together with Co minority $d$ states in the relevant energy window. These atom-centered states provide an efficient phase space for redistribution of the optically excited electronic population into local electronic states. The Co minority states also provide final states for the common ${\rm W}_{\downarrow}\rightarrow{\rm Co}_{\downarrow}$ OISTR-like channel, whereas the available W majority states support the additional ${\rm Co}_{\uparrow}\rightarrow{\rm W}_{\uparrow}$ channel. The resulting picture is consistent with the increase of the W local population, the development of W magnetization, and the persistent reduction of the interstitial population depicted in Fig. \ref{fig:FigIII}.

One of the most striking differences between Co/Pt and Co/W is the long-time behavior of their nonlocal electronic populations. Optical excitation initially drives electronic density into spatially extended states in both heterostructures. In Co/Pt this redistribution persists, leaving the interstitial population substantially above its initial value. In Co/W, however, the increase is transient. Following the initial excitation, the nonlocal population decreases and eventually becomes smaller than its equilibrium value, as expressed by Eq.~(\ref{eq:localization}). This is accompanied by increased occupation of atom-centered states, particularly states associated with the W layer. The effect therefore differs from a transient optical polarization of the electronic density during the pulse. We refer to this laser-induced transfer of electronic population from spatially nonlocal interstitial states toward atom-centered local states as \emph{light-induced electronic localization}. The PDOS provides an electronic-structure explanation for this behavior: Co/W offers substantial unoccupied $d$-orbital phase space within the energy window accessed by the optical excitation. This interpretation is consistent with previous observations of light-induced electronic localization in transition metals~\cite{Volkov2019,Vos2023}.

%-------------------------------------------------------------------------------------

\section{Discussion and Conclusion}

The comparison between Co/Pt and Co/W demonstrates that ultrafast spin-current
generation alone does not uniquely determine the subsequent electronic response
of an FM/NM heterostructure. Our previous calculations showed that ultrafast
longitudinal demagnetization generates an interlayer spin current whose temporal
profile approximately follows the magnetization rate
\cite{Kefayati2025Deciphering}. The present results reveal
another layer of microscopic physics: similar qualitative spin-current dynamics
can coexist with markedly different charge redistribution, spin-resolved
population dynamics, heavy-metal magnetization, and localization of the
photoexcited electronic population. These differences can be traced to the
material-dependent electronic phase space available to the optically excited
carriers.

An important result of the present calculations is the direct observation of interlayer electronic charge redistribution. This point is particularly relevant
in view of the nonlocal $sp$--$d$ exchange picture proposed for ultrafast spin transport in metallic multilayers, in which spin angular momentum can be
transferred from the FM to the NM through exchange coupling without an
accompanying transfer of charge carriers \cite{chen2023light}. Our calculations show a different
situation for Co/Pt and Co/W: the ultrafast spin dynamics coexist with a finite
redistribution of electronic charge between the two layers. Moreover, this
charge transfer is strongly material dependent. In Co/Pt, the initial
redistribution from Pt toward Co is relatively small and is subsequently
partially reversed, whereas Co/W exhibits a larger charge transfer
from Co toward W that persists on the simulated time scale. Thus, similar
qualitative interlayer spin-current dynamics do not imply similar interlayer
charge-transfer dynamics. The nonlocal-exchange and charge-transfer channels
should therefore not be regarded as universally mutually exclusive mechanisms;
their relative importance can depend on the electronic structure and available
phase space of the specific heterostructure.

The different temporal behavior of the charge and spin-resolved populations
provides further evidence that charge and spin redistribution cannot be
described as a single rigid transfer of carriers between the layers. The total
charge dynamics exhibit a delayed response of the FM layer relative to the NM, whereas the spin-resolved local populations display a delayed response of the NM relative to Co. This contrasting temporal ordering is natural because charge and spin observables probe different combinations of the two spin channels.
Consequently, redistribution between spin channels can produce a pronounced
spin response while giving a much smaller or delayed net charge response, and vice versa. The observed temporal offsets therefore point to the coexistence of several microscopic processes rather than to a single interlayer transport channel.

The spin-resolved populations provide additional information about these
processes. SOC-mediated spin redistribution and OISTR are present in both
heterostructures rather than defining mutually exclusive mechanisms for Co/Pt
and Co/W. In particular, a delayed minority-spin transfer from the NM toward Co
is observed in both systems and is consistent with an
${\rm NM}_{\downarrow}\rightarrow{\rm Co}_{\downarrow}$ OISTR channel into
available localized Co minority-spin states. Co/W additionally exhibits a much more pronounced majority-spin transfer from Co toward W. The distinction between Pt and W therefore lies not simply in the presence or absence of OISTR, but in the phase space available to the competing spin-resolved channels and hence in their relative magnitudes and directions.

For Co/Pt, the laser-accessible electronic phase space contains substantial
spatially extended $sp$-like character, consistent with the increasing
nonlocal electronic population and strong demagnetization of Co while Pt
remains nearly nonmagnetic. In Co/W, by contrast, the substantial available
atom-centered $d$-state phase space favors redistribution toward local states.
This behavior is accompanied by Co-to-W majority-spin transfer, persistent induced
magnetization of W, substantial interlayer charge redistribution, and a
persistent depletion of the interstitial electronic population. The
ground-state electronic structure therefore remains highly relevant even under
strongly nonequilibrium excitation because it defines the initial occupied
states and the available phase space into which the photoexcited electronic
population can evolve.

Another important distinction concerns the form of angular momentum transferred
between the layers. While we observe a net transfer of electronic spin angular
momentum, we do not find a corresponding net interlayer transfer of
\emph{electronic orbital angular momentum} between Co and Pt or between Co and
W \cite{Busch2023Sep,Seifert2023,Kumar2023nov,Kumar2023dec}. This observation
is relevant to interpretations of spintronic THz emission based on the
generation of an orbital current and its subsequent conversion into a charge
current through the inverse orbital Hall effect. For the Co/Pt and Co/W
heterostructures studied here, the interlayer electronic angular-momentum
transfer resolved in our calculations is predominantly associated with the
spin degree of freedom rather than with a net transfer of electronic orbital
angular momentum. We stress that this statement concerns the electronic
orbital angular momentum. Possible angular-momentum transfer involving lattice
degrees of freedom is not resolved here and is therefore outside the scope of
the present conclusion.

The interlayer charge redistribution also has a potentially measurable
electrodynamic consequence. The opposite changes of the electronic populations
of the FM and NM layers generate a transient charge separation across the
interface and therefore a time-dependent out-of-plane electric polarization.
Such a dynamical interfacial dipole can contribute to electromagnetic radiation
with a radiation geometry distinct from that associated with the conventional
in-plane transient charge current. We therefore suggest measurements of the
emitted electromagnetic field propagating along the plane of the FM/NM bilayer
as an experimental test of the predicted interlayer charge redistribution
\cite{Kefayati2024}. Such measurements could provide a complementary probe of
the out-of-plane charge dynamics that is not directly accessible from the
conventional THz-emission geometry.

Taken together, these results illustrate why interpreting spintronic THz emitters solely through a bulk spin Hall angle or a simple proportionality
between injected spin current and generated charge current can be incomplete. The microscopic response depends on the electronic structure of the complete FM/NM heterostructure and on the available spin-, orbital-, layer-, and spatially resolved phase space. This does not invalidate phenomenological descriptions such as superdiffusive transport, spin Hall conversion, or other effective transport pictures. Rather, the TDDFT results expose microscopic processes that such descriptions integrate out and identify the circumstances under which material-specific interface and electronic-structure effects become important.

Finally, the present calculations address the predominantly electronic regime during approximately the first 100~fs following optical excitation. At later times, coupling to lattice degrees of freedom and other relaxation channels can become increasingly important and may provide additional pathways for energy and angular-momentum transfer. Likewise, externally applied magnetic fields used in experiments can introduce additional collective spin dynamics \cite{Kefayati2026Light}. These
later-time processes lie beyond the scope of the present study.

\section{acknowledgments}
 This work was supported financially by Academia Sinica (AS), Taiwan (Grant No. AS-PD-1152-M03-2).

\bibliography{references}

\end{document}